\documentclass[aps,prl,twocolumn,superscriptaddress]{revtex4}
\usepackage{graphicx}

\begin{document}



\title{ Melting and metallization of silica in the cores of gas giants, ice giants and super Earths}

\author{S. Mazevet}
\affiliation{LUTH, UMR8102, Observatoire de Paris, CNRS, Universit\'e Paris Diderot, 5 place Jules Janssen, 92190 Meudon Cedex France}
\affiliation{CEA, DAM, DIF, F91297 Arpajon, France}
\author{T. Tsuchiya}

\affiliation{Geodynamics Research Center, Ehime University, 2-5 Bunkyo-cho, Matsuyama, Ehime 790-8577, Japan}
\author{T. Taniuchi}
\affiliation{Geodynamics Research Center, Ehime University, 2-5 Bunkyo-cho, Matsuyama, Ehime 790-8577, Japan}
\author{A. Benuzzi-Mounaix}
\affiliation{ LULI, Ecole Polytechnique, CNRS, CEA, UPMC, route de Saclay, 91128 Palaiseau, France} 
\affiliation{LUTH, UMR8102, Observatoire de Paris, CNRS, Universit\'e Paris Diderot, 5 place Jules Janssen, 92190 Meudon Cedex France}
\author{F. Guyot}
\affiliation{IMPMC, Muséum National d´Histoire Naturelle,, UMR  CNRS 75-90, Sorbonne Université Paris, France}


\begin{abstract} The physical state and properties of silicates at conditions encountered in the 
cores of gas giants, ice giants and of Earth like exoplanets now discovered with masses up to 
several times the mass of the Earth remains mostly unknown. Here, we report on theoretical 
predictions of the properties of silica, SiO$_2$, up to 4 TPa and about 20,000K using first 
principle molecular dynamics simulations based on density functional theory. For conditions found 
in the Super-Earths and in ice giants, we show that silica remains a poor electrical conductor 
up to 10 Mbar due to an increase in the Si-O coordination with pressure.  For Jupiter and Saturn 
cores, we find that MgSiO$_3$ silicate has not only dissociated into MgO and SiO$_2$, as shown in previous 
studies, but that these two phases have likely differentiated to lead to a core made of liquid SiO$_2$ and 
solid (Mg,Fe)O.  \end{abstract}

\maketitle

Little is known of the properties of silicates at pressures greater than those encountered 
in the Earth's mantle \cite{baraffe}. Beyond 100 GPa, and up to the conditions encountered in Jupiter’s 
interior, 4-7 TPa, our current understanding is limited to extrapolations of equations of state 
of the solid phases only valid at moderately low temperatures (few thousands Kelvins) \cite{umemoto}.
On the experimental side, the high-pressures (up to ≈ 1TPa) and high-temperatures
(up to 50 000 K) domain has been investigated along the principal Hugoniot using dynamical 
experiments \cite{hicks,knudson,spaulding}. Dynamical compression techniques based on quasi-isentropic path to 
reduce heating are in progress, but experimental data for super Earth and Jupiter conditions will remain sparse 
in the near furture. For years to come, the modeling of planetary interiors will mostly rely on calculations,
such as first principles simulations, and in the optimum case, validated on the restricted pressure domain
accessible experimentally. 
In the meantime, the discovery of extrasolar planets several times the mass of the Earth such 
as CoRo 7b (3 Earth mass) \cite{rivera,qeloz} or the modeling of the interior structure and evolution 
of Jupiter, Saturn, Uranus and Neptune in our solar system all require to elucidate the phase diagram 
of silicates well beyond a few Mbar\cite{baraffe}. 

According to accepted planetary formation scenario\cite{guillot}, silicates are, together with water 
and iron, the main ingredients of the proto-planet embryos. They can grow up to several times the 
size of the Earth and then accrete hydrogen and helium up to several times the mass of Jupiter or 
Saturn. The properties of silicates throughout this trajectory in pressure and temperature space 
need to be better constrained. If they turn liquid, it potentially impacts dissipation and tidal 
effects that are keys to understand the stability of planetary systems, the erosion of the core of 
giant planets or even their luminosity if differentiation leads to a significant release of heat. 
If they turn metallic in pressure, silicates can significantly contribute to the generation of 
magnetic fields or play a major role in dissipation effects in terrestrial exoplanets. 

  In contrast to the iron and water phase diagrams that are better constrained in this 
regime\cite{redmer,stixrude,tsuchiya}, the complex behavior of silicates in density and temperature 
is currently a challenge for first principle simulations based on Density Functional Theory (DFT). 
Complex silicates constituting planetary interiors such as (Mg,Fe)SiO$_3$ and (Mg,Fe)$_2$SiO$_4$ 
cannot be directly simulated due to the large numbers of particles needed. A first approach to the 
stability and complex chemistry occurring at pressures and temperatures relevant to planetary 
interiors can be inferred by studying the properties of compounds such as MgSiO$_3$, a major silicate 
endmember in the deep Earth interior and its dissociation products MgO, and SiO$_2$ \cite{umemoto}.

\begin{figure}
\includegraphics[scale=0.6]{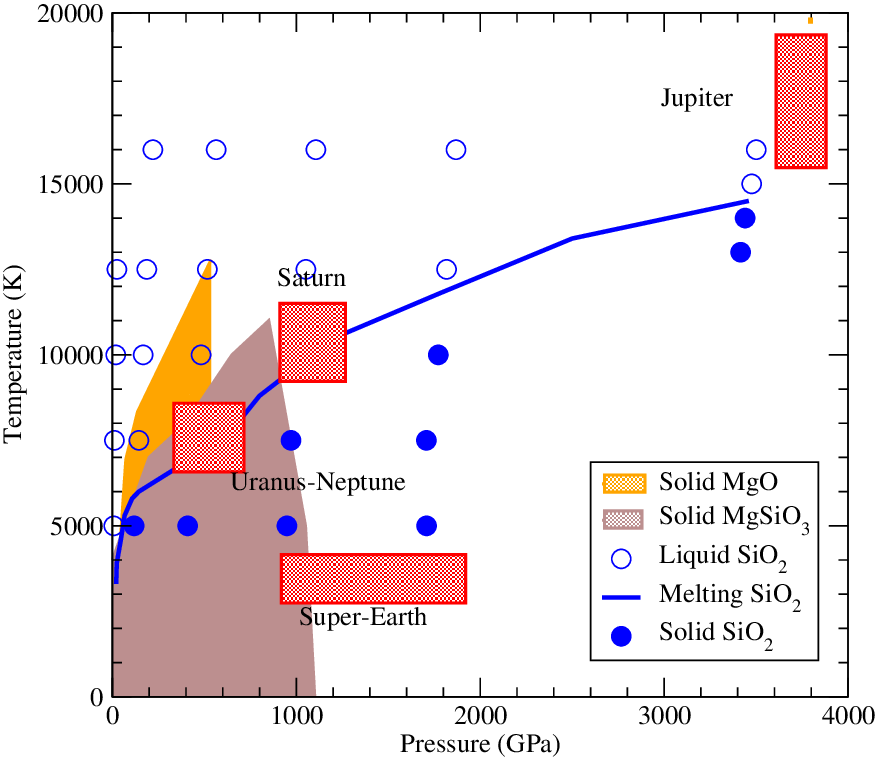}
\vspace{0.5cm}\caption{Phase diagram of MgSiO$_3$, SiO$_2$, MgO. (Orange) stability of MgSiO$_3$ in pressure and temperature. (Brown) Solid MgO. (Blue) SiO$_2$ high-pressure melting curve. (Filled blue dots) calculated solid SiO2, (open blue dots) calculated liquid SiO2}\label{fig1}
\end{figure}

Using lattice dynamics, Umemoto et al.\cite{umemoto} established the stability of MgSiO$_3$ (Fig.\ref{fig1}) 
up to pressures representative of Jupiter's interior. 
Within the quasi-harmonic approximation, they showed that CaIrO$_3$ –type MgSiO$_3$, the silicate stable 
at the terrestrial core-mantle boundary, cannot exist beyond 1 TPa and probably melts at 
temperatures varying between 5000K and 10,000K in this pressure range. As shown in Fig.\ref{fig1}, this 
implies that MgSiO$_3$ may remain stable at conditions encountered in Uranus and Neptune but that 
it would dissociate into MgO and SiO$_2$ in Saturn and Jupiter. As the quasi-harmonic approximation tends to
overestimate the melting temperature, whether MgSiO$_3$ is in a liquid or a solid state at conditions
found in Uranus and Neptune cores remains an open question. Its answer necessitates to go beyond the 
quasi-harmonic approximation by including anharmonic effects using molecular dynamics. As a first step 
toward a full ab initio phase diagram of silicates in the pressure-temperature range 
relevant to planetary modeling, we calculated the physical properties of silica, SiO$_2$, 
using DFT based molecular dynamics simulations (DFT-MD) for the high pressure melting 
curve and linear response theory for the electrical properties.

For the high-pressure melting curve, 
we performed two-phase molecular dynamics simulations up to 4 TPa as this pressure range is 
representative of Jupiter's interior. We used specifically designed pseudo-potentials with 2s and 2p 
shells included as valence states to reach the highest pressures investigated here. This suit of
pseudo-potentials were validated against all-electron calculations for the relevant range of
pressure\cite{jollet} Our two-phase 
MD simulations were performed using the pwscf code \cite{giannozzi} for electronic structure with 
an original implementation of the temperature constant MD solver, where the temperature $T$ was 
controlled to keep constant values by the kinetic energy scaling method\cite{usui}. 

Two phases, both with the melt and crystalline structures, were placed in a computational box. Here, 
both phases contained 24 SiO$_2$ units (72 atoms), and the computational box contained 48 units (144 atoms) 
total. 
The simulation procedure was as follows: the crystalline structure at each pressure was first 
equilibrated at 2 000 K for a period of 1 ps (10$^{-12}$ s). The liquid phase at each pressure were 
obtained by equilibrating at 20 000 K for a period of 1 ps. The solid and liquid structures were put 
together with a small spacing in a simulation box, letting the lower part of the box be solid while 
keeping the upper part liquid. We thermalized the liquid cells for 50 time steps at target temperatures 
before starting two phases- simulations. In our simulations, the solid and liquid cells were connected 
at the (001) plane of solid. Starting from these configurations, the MD simulations evolve into a single 
phase. For temperatures above the melting temperature, $T_m$, the cell will turn liquid, whereas the cell 
will solidify in a vitreous state at temperature below $T_m$. The simulations were typically performed 
for 10 000 time steps (10 ps). $T_m$ at pressures below 160 GPa were obtained from reference\cite{usui}.
\begin{figure}
\includegraphics[scale=0.6]{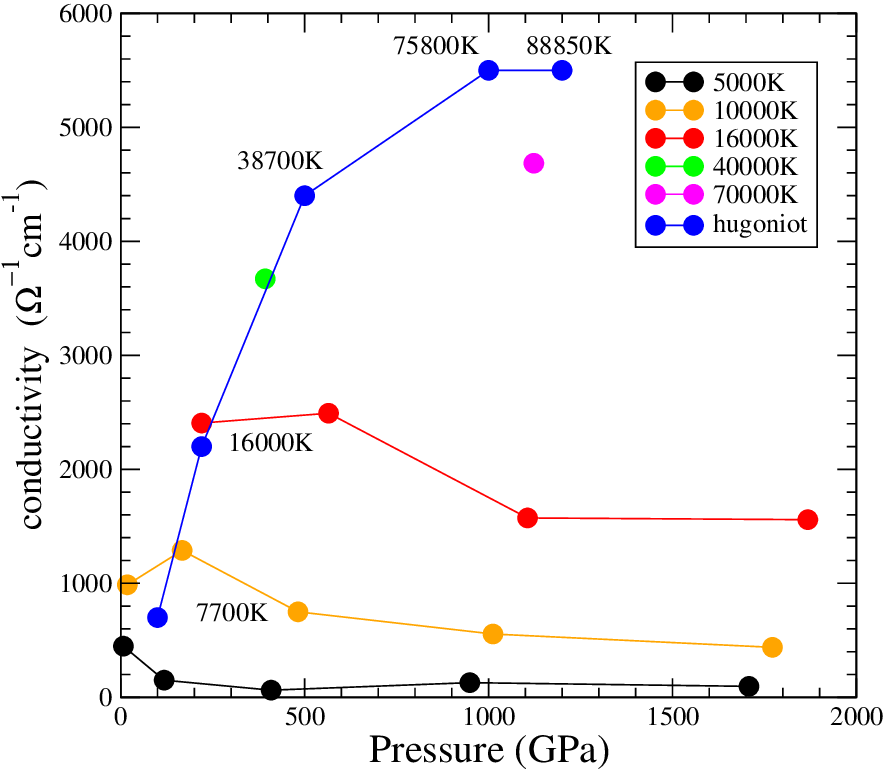}
\vspace{0.5cm}\caption{Variation of the electrical conductivity in pressure along various isotherms and at conditions relevant to planetary modeling. Solid lines are guides to the eyes.}\label{fig2}
\end{figure}

In the SiO$_2$ compound, several phases have been identified as stable at ultra-high-pressures including an 
hexagonal Fe$_2$P-type structure at 1 TPa, beyond a cotunnite-type phase \cite{tsuchiya}. These solid 
phases were used in two-phases simulations to calculate the melting temperature. This led to 
different slopes in the melting curve and to discontinuities observable in Fig.\ref{fig1}. From 0.1TPa and up 
to 1 TPa, Fig.\ref{fig1} shows that the melting temperature increases up to 10,000K. This region corresponds 
to conditions anticipated at the cores of Uranus, Neptune, and Saturn \cite{baraffe}. 

Fig.\ref{fig1} also shows that for these three planets, a core constituted of pure silica would be molten 
in the core as the pressure and temperature conditions anticipated lay above the calculated high pressure-melting 
curve. The situation is somewhat tempered in the cases of Uranus and Neptune as the quasi-harmonic 
approximation estimates that CalrO$_3$-type MgSiO$_3$ remains stable at these conditions\cite{umemoto}. 
As anharmonic effects may lead to a lower melting temperature and a smaller MgSiO$_3$ stability domain 
of solid phases in pressure and temperature, the calculations shown here suggest that the cores of 
Uranus and Neptune-type planets would also be constituted of a significant fraction of melted 
silica if MgSiO$_3$ is shown to be dissociated at these conditions. 

The situation for Jupiter and Saturn's cores is clearer. Fig.\ref{fig1} shows that silica is molten at these 
conditions. In this pressure-temperature range, CalrO$_3$-type MgSiO$_3$ is not stable and is 
predicted to dissociate into MgO and SiO$_2$ within the quasi-harmonic approximation. We also note 
that DFT-MD simulations \cite{belonoshko} show that the MgO melting temperature increases with 
pressure much faster than either those of MgSiO$_3$ or SiO$_2$ up to 0.5 TPa. An extrapolation of the MgO 
high-pressure melting curve up to the conditions of Saturn's core, 1TPa, suggests that it remains 
solid and probably in the B2 phase. This indicates that during its evolution, Saturn's silicates 
are probably dissociated into two components, solid MgO and liquid SiO$_2$. Considering the rather 
close densities of these two constituents at a pressure of 0.5 TPa (7 g/cm3 and 7.2g/cm3 respectively)
and without direct calculations of the MgO properties at 1TPa, the calculations shown here indicate 
that Saturn's silicates could turn in either a MgO shell surrounding a silicate liquid core or conversely 
a MgO core surrounded by a layer of liquid silica. The presence of iron, likely preferentially 
dissolved in MgO would increase the density of this latter phase and favor a core made of a central 
(Mg,Fe)O surrounded by molten SiO$_2$.  These considerations also apply to Jupiter’s core. We note, 
however, that it exists at the moment a wide disparity between models regarding the conditions at 
Jupiter’s core with predictions for temperatures at the envelopp-core boundary ranging from 
15,000 to 20,000K \cite{baraffe}.

Fig.\ref{fig1} also shows that silica remains in a solid state in Earth like exoplanets up to several times 
the Earth mass. Whether silica turns metallic at these conditions is a second important issue that 
we investigated using linear response theory.  To obtain the transport properties, we performed a set of 
simulations starting from the quartz alpha phase with 108 atoms in the simulation cell. We used the isokinetic 
ensemble and performed the calculations with the Abinit electronic structure code\cite{gonze}. We checked 
that the simulation relaxes into the proper phase at a given pressure-temperature condition by calculating 
the Si-O coordination number. From equilibrated trajectories, we performed Kubo-Greenwood calculations 
on several snapshots to obtain the transport properties. Within this formulation\cite{mazevet}, the 
real part of the optical conductivity is obtained by using the eigenvalues and eigenfunctions obtained from the 
diagonalization of the Kohn-Sham Hamiltonian.  We used 5 snapshots and a 43 k-points grid as given by the 
Monkhorst-Pack scheme \cite{monkhorst} to insure convergence of the optical conductivity in particle numbers, 
k-points, and number of snapshots. We used 800 bands to insure convergence of the optical conductivity in 
photon frequency well beyond 10eV. The direct current (DC) conductivity plotted in Fig.\ref{fig2} is obtained 
by taking the limit of the optical conductivity at zero frequency. 
\vspace{1cm}\begin{figure}
\includegraphics[scale=0.6]{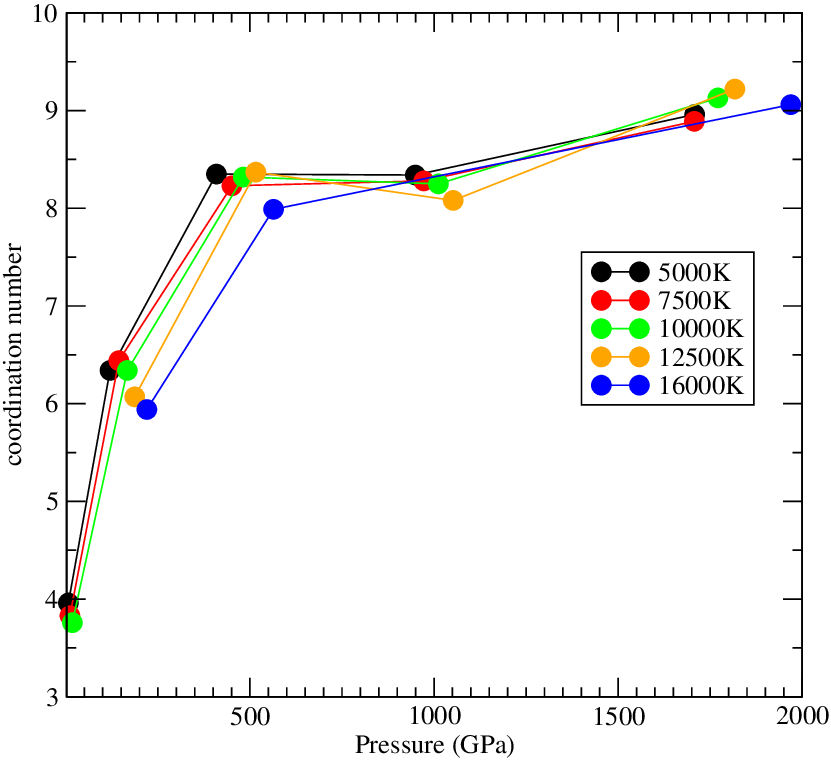}
\vspace{1cm}\caption{Variation of the coordination number (CON) in pressure along five isotherms. }\label{fig3}
\end{figure}

Fig.\ref{fig2} shows DC conductivity calculated using supercells equilibrated along 
several isotherms as indicated in Fig.\ref{fig1}. We find that the DC conductivity increases as temperature 
increases along the principal Hugoniot (i.e. the final states reached in a shock experiment starting from 
normal conditions) to reach a plateau at 5500 ohm$^{-1}$cm$^{-1}$, concomitantly with the disappearance of 
Si-O bonding \cite{laudernet}. This is in very good agreement with shock wave experiments\cite{hicks}. 
Surprisingly, Fig.\ref{fig2} also shows that no non-metal-metal transition is observed in silica as pressures 
increase along an isotherm. The DC electrical conductivity actually does not increase with pressure 
along a given isotherm but rather decreases. This is unexpected since many insulators undergo a 
pressure-induced metallization at relatively low temperatures \cite{mott,eremets}. The origin of this 
effect can be traced back to the change in the Si-O coordination with pressure and the corresponding 
variation of the electronic structure. 

It is well documented that the Si-O coordination increases from 4 in alpha-quartz to 9 in the cotunnite and 
Fe$_2$P phases\cite{tsuchiya}. Fig.\ref{fig3} shows the values of the coordination number (CON) obtained by 
integrating the Si-O correlation function up to $r_{cut}=4.5$a$_B$ at the lowest densities and up to 
$r_{cut}=4$a$_B$ at 
9g/cm$^3$ and 11g/cm$^3$. These cutoff values correspond to the minimum in the Si-O pair correlation function 
following the first peak. Fig.\ref{fig3} shows that the Si-O CON has a similar behavior in either the solid or 
the liquid at these temperatures with values corresponding to the ones obtained previously for the Fe$_2$P-type 
structure \cite{tsuchiya}. Fig.\ref{fig3} also shows that temperature has little effect on the Si-O CON at these 
conditions, although a decrease in the Si-O CON in temperature is still noticeable at each density. The increase 
in the Si-O CON with pressure corresponds to a drastic modification of the electronic structure. 
\vspace{1cm}\begin{figure}
\includegraphics[scale=0.6]{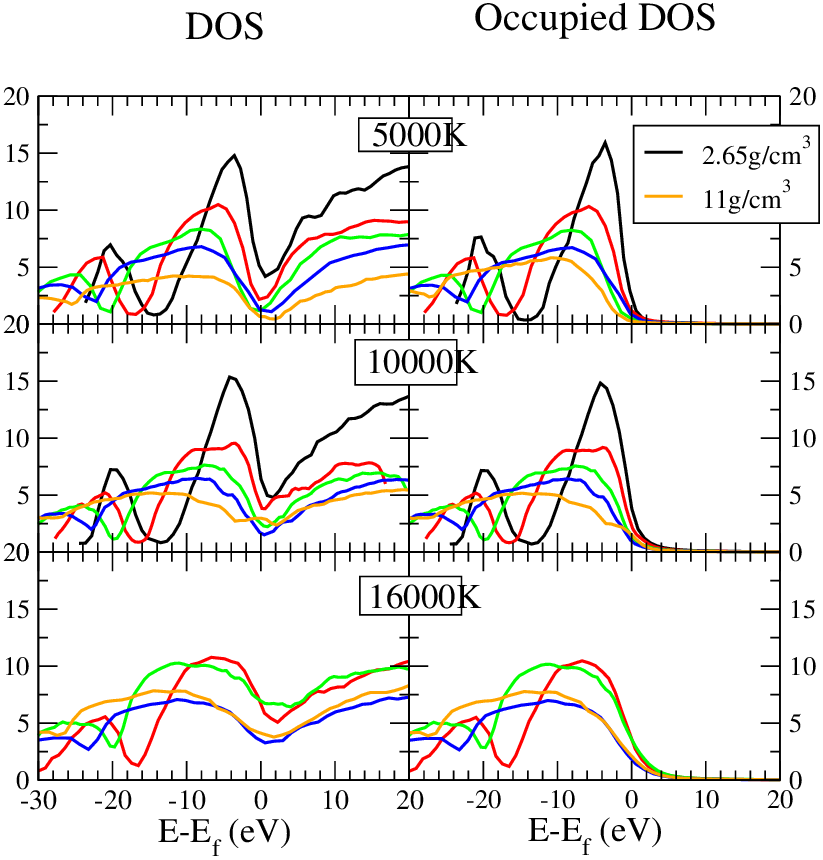}
\vspace{1cm}\caption{Variation of the density of states (DOS) and occupied density of states in temperature and pressure. Each raw corresponds to an isotherm as indicated in the figure. The densities are (black) 2.65g/cm$^3$, (red) 5g/cm$^3$, (green) 7g/cm$^3$, (blue) 9g/cm$^3$, (orange) 11g/cm$^3$.}\label{fig4}
\end{figure}

Fig.\ref{fig4} shows  the density of states (DOS) and occupied density of 
states (occupied DOS) obtained at the densities corresponding to the conductivities given in Fig.\ref{fig2} 
and for three representative isotherms. Fig.\ref{fig4} shows that silica behaves as a semi-metal at low density 
and moderate temperatures with a pseudo-gap clearly visible around the Fermi energy. Along an isotherm, the 
increase in CON corresponds to the opening of a pseudo-gap and a decrease in the density 
of state around the Fermi energy. The latter is clearly visible in the occupied DOS. Conversely, comparing the 
DOS and occupied DOS in between isotherms at a given density shows that temperature has the opposite effect on 
the electronic structure. This suggests that pressure along an isotherm tends to increase the localization of 
the electron on each atom of the Si-O bond, i.e. implicitly increasing its ionic character.  
It also shows that the effect of temperature is more involved than a simple 
temperature activated process across a gap fixed at a given density. As temperature decreases the Si-O CON 
increases at a given density, the DOS and occupied DOS are modified accordingly and lead to smaller 
conductivities as shown in Fig.\ref{fig2}. 

This behavior of silica leads to a significant correction of previous estimates of the 
electrical conductivity by an order of magnitude\cite{umemoto}. We find that the DC 
electrical conductivity at conditions relevant to the deepest mantles of super-Earths is of the 
order of 200-300 ohm$ˆ{-1}$cm$ˆ{-1}$. Contrary to previous speculations, this suggests that silica will not 
contribute significantly to either magnetic field generation in Earth like exoplanets up to 
several times the Earth mass. The DC electrical conductivity of silica starts to be comparable 
to the one of a poor metal (greater than 1000 ohm$^{-1}$cm$^{-1}$) above 7000-8000K. 

We established here new equation of state data for silica needed to build self-consistent planetary 
models of ice giants, gas giants, and terrestrial exoplanets . 
Contrary to previous estimates, we find that silica remains a poor 
metal at extreme pressures with little contribution to transport properties and magnetic field 
in terrestrial exoplanets. The molten state of SiO2 at conditions deep in Jupiter and Saturn, and possibly 
in Uranus and Neptune, suggests that structure and evolution models of such bodies may need to account 
for a liquid state in their cores.  This is also the case for standard tidal models that consider 
dissipation as mostly occurring in a solid core made of silicates, water and iron\cite{remus}.

\begin{acknowledgments}
 This work is supported in part by the 
French Agence National de la Recherche under contract PLANETLAB ANR-12-BS04-0015.
\end{acknowledgments}

\end{document}